\documentclass[11pt]{article}

\usepackage[margin=0.93in]{geometry}
\usepackage{amsmath,amsthm,amssymb}
\usepackage{graphicx}
\usepackage{multirow}
\usepackage{paralist}

\newtheorem{theorem}{Theorem}

\newtheorem{corollary}[theorem]{Corollary}

\newcommand{\N}{\mathbb{N}}
\newcommand{\BO}{\mathcal{O}}
\newcommand{\E}{\mathbb{E}}

\usepackage[colorlinks,linkcolor=blue,filecolor=blue,citecolor=blue,urlcolor=blue,pdfstartview=FitH,hypertexnames=false]{hyperref}

\newcommand{\namedref}[2]{\hyperref[#2]{#1~\ref*{#2}}}
\newcommand{\sectionref}[1]{\namedref{Section}{#1}}

\newcommand{\tableref}[1]{\namedref{Table}{#1}}

\newcommand{\equalityref}[1]{\hyperref[#1]{Equality~\eqref{#1}}}
\newcommand{\inequalityref}[1]{\hyperref[#1]{Inequality~\eqref{#1}}}

\DeclareMathOperator{\polylog}{polylog}

\begin{document}

\setcounter{page}{0}

\setcounter{tocdepth}{3}

\title{The 1-2-3-Toolkit for Building Your Own Balls-into-Bins Algorithm}

\author{
Pierre Bertrand\thanks{Ecole Normale Superieure Cachan. \hbox{Email: {\tt
pierre.bertrand@ens-cachan.fr}.}}
\and
Christoph Lenzen\thanks{MPI for Informatics, Saarbr\"ucken, Germany.
\hbox{Email: {\tt clenzen@mpi-inf.mpg.de}.}
Supported by the Deutsche Forschungsgemeinschaft (DFG, reference
number Le 3107/1-1).}
}

\date{}

\maketitle

\begin{abstract}
In this work, we examine a generic class of simple distributed balls-into-bins
algorithms. Exploiting the strong concentration bounds that apply to
balls-into-bins games, we provide an iterative method to compute accurate
estimates of the remaining balls and the load distribution after each round.
Each algorithm is classified by (i) the load that bins accept in a given round,
(ii) the number of messages each ball sends in a given round, and (iii) whether
each such message is given a \emph{rank} expressing the sender's inclination to
commit to the receiving bin (if feasible). This novel ranking mechanism results
in notable improvements, in particular in the number of balls that may commit to
a bin in the first round of the algorithm. Simulations independently verify the
correctness of the results and confirm that our approximation is highly accurate
even for a moderate number of $10^6$ balls and bins.
\end{abstract}

\section{Introduction \& Related Work}

Consider a distributed system of $n$ anonymous balls and $n$ anonymous bins,
each having access to (perfect) randomization. Communication proceeds in
synchronous rounds, each of which consists of the following steps.
\begin{compactenum}
  \item Balls perform computations and send messages to bins.
  \item Bins receive messages, perform computations, and respond to the received
  messages.
  \item Each ball may commit to a bin, inform it, and terminate.\footnote{Observe
  that this step can be safely merged with the first step of the subsequent
  round. Hence, the communication delay incurred by an $r$-round algorithm
  equals that of $r$ round trips plus the one of a final commit message.}
\end{compactenum}
The main goals are to minimize the maximal number of balls committing to the
same bin, the number of communication rounds, and the number of messages. This
fundamental load balancing problem has a wide range of applications, including
job assignment tasks, scheduling, low-congestion routing, and hashing,
cf.~\cite{mitzenmacher01}.

The first distributed formulation of the problem was given in
1995~\cite{adler95}. Among other things, in this work it was shown that even a
single round of communication permits an exponential reduction of the bin load
compared to the trivial solution of each ball committing to a random bin,
without increasing the number of messages bins and balls send by more than a
constant factor. In the sequel, a number of publications established a clear
picture of the asymptotics of the
problem~\cite{even09,even10,lenzen11tight,stemann96}.
Algorithms that run for $r$ rounds and are \emph{non-adaptive}---each ball
chooses the bins it communicates with in advance---and
\emph{symmetric}---contacted bins are chosen uniformly and independently at
random (u.i.r.)---can obtain maximal bin load $\Theta((\log n/\log \log
n)^{1/r})$~\cite{even10,stemann96}.\footnote{The lower bound applies for
constant values of $r$ and the number of contacted bins only.} Symmetric
algorithms sending $\BO(n)$ messages in total require at least
$(1-o(1))\log^*n-\log^*L$ rounds to achieve bin load $L$, while $\log^*
n+\BO(1)$ rounds are sufficient for bin load $2$~\cite{lenzen11tight}. Finally,
without such constraints, a maximal bin load of $3$ can be guaranteed within
$\BO(1)$ rounds~\cite{lenzen11tight}.

Unfortunately, this information is of little help to a programmer or system
designer in need of a distributed balls-into-bins subroutine. How should one
decide which algorithm to pick? For a reasonable value of $n$, say $10^5$, the
constants in the above bounds are decisive. For instance, $\log 10^5/\log \log
10^5\approx 4$, whereas $\log^* 10^5=5$. Moreover, some bounds are not very
precise. For example, Stemann proves no results for maximal bin loads smaller
than $32$~\cite{stemann96}. This is a constant, but arguably of little practical
relevance: For $n=10^9$, letting each ball commit to a random bin results in
maximal bin load smaller than $16$ with probability larger than $99.98\%$. Yet,
our simulations show that Stemann's algorithm performs better than an adaptive
variant of the multi-round Greedy algorithm, for which loads of 3-4 in 3 rounds
are reported for $n\in[10^6,8\cdot 10^6]$~\cite{even09}. The symmetric algorithm
from~\cite{lenzen11tight} guarantees an even better bin load of $2$, but, again,
the asymptotic round complexity bound of $\log^* n+\BO(1)$ seems overly
cautious: the corresponding lower bound basically just shows that a single round
is insufficient to this end.

In summary, the existing results are inconclusive for relevant parameter ranges:
values of $n$ that may occur in practice admit very few rounds and small loads,
even for symmetric algorithms. Hence, it seems natural to explore this
region, aiming for accurate estimates and small loads.

\textbf{Contribution.} We analyze two types of simple symmetric algorithms. The
first class subsumes the symmetric algorithm from~\cite{lenzen11tight}. The
second, novel class strictly improves on the first in terms of the number of
balls that can be placed with the same maximal bin load, number of rounds, and
message complexity. In each round $i\in\N$, our algorithms perform the following
steps.
\begin{compactenum}
   \item Each ball sends a number $M_i\in \N$ of messages to uniformly
   independently random (u.i.r.) bins. These messages are either identical
   or are ranked $1,\ldots,M_i$.
   \item A bin of current load $\ell$ responds to (up to) $L_i-\ell$ balls,
   where smaller ranks are preferred. Ties are broken by choosing uniformly at
   random.
   \item Each ball that receives at least one response commits, either to a
   random responding bin (for unranked messages), or the responding bin to
   which it sent the message of smallest rank.
\end{compactenum}
We further restrict that $L_{i+1}\geq L_i$ for all $i\in \N$, since there is
little use in decreasing the accepted loads in a later round. An algorithm is
thus described by the sequence $(M_i)_{i\in \N}$, the increasing sequence
$(L_i)_{i\in \N}$, and whether messages are ranked or not.

We provide an analytical iterative method for computing accurate estimates of
the number of committed balls and the load distribution after each round. While
no technical innovation is required to this end, finding accurate and simple
expressions for the involved expectations proved challenging. Moreover, we
devised a program that, given the above parameters, computes these values. Our
approach extends to the general case where there is a different number of balls
and bins. We complement our analysis by simulations, which serve to double-check
the correctness of the analytical bounds and confirm that they are highly
accurate for practical values of $n$. Furthermore, we compare to the algorithms
from the literature by means of simulation.\footnote{The code is available
online~\cite{scripts}.}

\textbf{Main Results.} The derived bounds (confirmed by our simulations) show
that symmetric algorithms can achieve bin loads of $2$ or $3$ within $2$ to $3$
rounds, using fewer than $6n$ messages.\footnote{Note that sending fewer than
$3n$ messages implies that some balls commit without receiving any message.}
Since we allow for arbitrary sequences $(M_i)_{i\in \N}$, we can also infer what
can be achieved if the number of messages balls sent in each round is capped at
a small value. For instance, with $M_1=1$ and $M_2=M_3=2$, $n=10^6$ balls can
reliably placed within $3$ rounds, with a maximal bin load of $3$ and fewer than
$3.5n$ messages in total. Here it proves useful to pick $L_1=2$ and increase the
permitted load to $3$ only in later rounds, ensuring that balls can be placed
reliably despite sending few messages. For all the choices of parameters we
considered, previous algorithms are consistently outperformed by our approach.

Due to the variety of parameters, algorithms, and optimization criteria, it is
difficult to provide a general answer to the question which algorithm to use
(with which parameters). Therefore, we consider the method by which we derive
our bounds and the program code permitting their fast and simple evaluation to
be of independent interest. In a practical setting, we expect that the available
knowledge on constraints and optimization criteria will make the search space
sufficiently small to tailor solutions with good performance using the toolbox
we provide.

\textbf{Paper Organization.} In \sectionref{sec:preliminaries}, we discuss why
all random variables of interest are highly concentrated around their
expectations and introduce notational conventions. In \sectionref{sec:first}, we
analyze the first round of our algorithms. We discuss how to extend the approach
inductively to rounds $2,3,\ldots$ in \sectionref{sec:later}, as well as how to
apply it to the general case of $m\neq n$ balls. In \sectionref{sec:specific},
we evaluate a few choice sets of parameters to shed light on the performance of
the resulting algorithms and, by means of simulation, compare to algorithms from
the literature. Finally, in \sectionref{sec:conclusion} we draw some
conclusions.

\section{Preliminaries}\label{sec:preliminaries}

\textbf{Concentration Bounds.}
For the considered family of algorithms, sets of random variables like whether
bins receive at least $m\in \N$ messages in a given round are not independent.
However, they are \emph{negatively associated}, implying that Chernoff's bound
is applicable \cite{dubhashi96}. Denoting for any constant $m$ by
$X_{\geq m}$ ($X_m$) the number of bins receiving at least (exactly) $m$
messages in round $1$, it follows for any $\delta>0$ that
\begin{align*}
{}& P(|X_m-\E[X_m]|>\delta \E[X_m])\\
\leq {}&   P\left(|X_{\geq m}-\E[X_{\geq m}]|>\frac{\delta \E[X_{\geq
m}]}{2}\right) + P\left(|X_{\geq m+1}-\E[X_{\geq m+1}]|>\frac{\delta \E[X_{\geq
m}]}{2}\right)\\
\leq {}&  P\left(|X_{\geq m}-\E[X_{\geq m}]|>\frac{\delta \E[X_{\geq
m}]}{2}\right) + P\left(|X_{\geq m+1}-\E[X_{\geq m+1}]|>\frac{\delta \E[X_{\geq
m+1}]}{2}\right)\\
\leq {}& 4e^{-\delta^2\min\{\E[X_{\geq m}],\E[X_{\geq m+1}]\}/16},
\end{align*}
where we applied Chernoff's bound to each of the random variables in the last
step. Note that (i) trivially $\E[X_{m}]\in O(n)$ and (ii) $\E[X_{m}]$ decreases
exponentially in $m$ for $m\geq M_1\in \BO(1)$, the expected number of messages
a bin receives in round $1$, as messages are sent to u.i.r.\ bins. Hence, for
any natural number $\gamma$,
\begin{align*}
{}& P\left(\sum_{m\in \N_0}|X_m-\E[X_m]|\leq \gamma^3\sqrt{n}\right)\\
 \geq {}&  
1-\sum_{m=0}^{\gamma^2-1}P\left(|X_m-\E[X_m]|> \gamma\sqrt{n}\right)
-\sum_{m=\gamma^2}^{M_1n}P[X_m>0] 
\tag*{union bound}\\
\in {}& 
1-\gamma^2e^{-\Omega(\gamma^2)}-\sum_{m>\gamma^2}^{M_1n}n\cdot
e^{-\Omega(m/M_1)} \tag*{Chernoff (left), Markov + union bound (right)}\\
\subseteq {}& 1-e^{-\Omega(\gamma^2)+\BO(\log n)}.\tag*{$M_1$ constant}
\end{align*}
Put simply, assuming that the random variables $X_m$ attain their expected value
in all computations introduces only a marginal error: The probability that, say,
more than $\sqrt{n}\log^3 n$ bins receive a different number of messages than
they would if we just ``assigned'' messages according to expecations is at most
$n^{-\Omega(\log n)}$. Similar reasoning applies in case the algorithm utilizes
ranks.

\begin{corollary}[follows from \cite{dubhashi96} as shown
above]\label{coro:concentration}
For any $\gamma\in \N$,
\begin{equation*}
P\left(\sum_{m\in \N_0}|X_m-\E[X_m]|\leq \gamma^3\sqrt{n}\right)\in
1-e^{-\Omega(\gamma^2)+\BO(\log n)}.
\end{equation*}
\end{corollary}

Once the message distribution is fixed, bins decide to which balls to respond.
Since ties are broken by u.i.r.\ choices of the bins, the results from
\cite{dubhashi96} show that the number of balls receiving a certain number of
responses also obey Chernoff's bound. Finally, we conclude that the random
variables counting the number of bins with a given load after the first round is
subject to Chernoff's bound as well. In summary, all the variables we will
consider are tightly concentrated.

By induction, this reasoning extends to (a constant number of) subsequent
rounds. When the total number of sent messages becomes smaller, also the
deviation from the expected values we need to consider becomes smaller (i.e., we
can replace the factor $\sqrt{n}$ above by the root of the largest considered
expected value); concentration for the number of bins receiving no message
follows from the bounds for the other variables. Overall, these considerations
imply that for the purposes of this work, it is sufficient to assume that the
aforementioned random variables match their expectation, as the induced error is
negligible. Simulations will confirm this view; for the sake of a
straightforward presentation, we hence refrain from phrasing statements
analogous to Corollary~\ref{coro:concentration} for the random variables
considered throughout this paper.

\textbf{Notational Convention.}
Given the above observations, we will base our analysis on expected values. This
entails that we implicitly neglect terms of lower order, and it will be
convenient to do so when computing probabilities as well. For instance, cleary
\begin{equation*}
\frac{\binom{n}{k}}{n^k/k!}=\prod_{i=1}^{k}\frac{n-(i-1)}{n}
\geq \left(1-\frac{k}{n}\right)^k\geq 1-\frac{k^2}{n}.
\end{equation*}
Using the approximation $\binom{n}{k}\approx n^k/k!$ for $k\in \BO(n^{1/4})$
when computing a probability will thus not incur a total error of more than
$\BO(\sqrt{n})$ when infering an expectation. Therefore, we adopt the convention
of writing $x\approx y$ whenever $x\in (1\pm \polylog n / \sqrt{n})y$ (for
probabilities or expectations).

\section{The First Round}\label{sec:first}

\subsection{Unranked Messages}

In order to compute the (approximate) probability $p$ that a ball successfully
commits in the first round, we need to determine how likely it is to receive a
response to a message from a bin. To this end, we let all balls but one make
their random choices and determine the expected number $\E[X_m]$ of bins with
$m$ messages. As argued in \sectionref{sec:preliminaries}, the $X_m$ are sharply
concentrated around their expectation, so this is sufficient for estimating $p$
with negligible error. Note also that, choosing $\gamma\in \Omega(\sqrt{\log
n})$, we can apply the union bound over all $n$ balls to see that this estimate
is accurate for \emph{all} balls concurrently.

Since $M_1n$ messages are sent to u.i.r.\ bins in the first round, we have
\begin{equation*}
\E[X_m]= n\cdot \binom{M_1n}{m}\left(\frac{1}{n}\right)^m
\left(1-\frac{1}{n}\right)^{M_1n-m}
\approx  n\cdot\frac{M_1^{m}}{m!}
\left(1-\frac{1}{n}\right)^{M_1n}
\approx  n\cdot \frac{M_1^m}{e^{M_1}m!}.
\end{equation*}
Recall that each bin chooses a subset of at most $L_1$ received messages to
respond to. The probability that a ball may commit is thus
\begin{equation}\label{eq:p_round_1}
p(M_1,L_1)\approx 1-(1-p_s(M_1,L_1))^{M_1},
\end{equation}
where $p_s(M_1,L_1)$ is the probability that a single message does result in a
response. Note that we ``held back'' the messages of the ball in question when
approximating the number of bins with a given load. Hence we need to add one to
the load of a contacted bin when determining the probability that it responds to
a message. We compute
\begin{equation*}
p_s(M_1,L_1)\approx \sum_{m=0}^{L_1-2}\frac{M_1^m}{e^{M_1}m!} +
\sum_{m=L_1-1}^{\infty}
\frac{M_1^m}{e^{M_1}m!}\cdot \frac{L_1}{m+1}
=\frac{1}{e^{M_1}}\sum_{m=0}^{L_1-2}\frac{M_1^m}{m!} + 
\frac{L_1}{M_1e^{M_1}}\sum_{m=L_1}^{\infty}\frac{M_1^m}{m!}.
\end{equation*}
Inserting these values into \equalityref{eq:p_round_1}, we obtain the
(asymptotic) percentage of balls that will not commit in the first round, given
in \tableref{table:round_1} in the appendix; simulation results from $100$ runs,
each with $10^6$ balls and bins, confirm the tight concentration of the values.

We see that increasing the number of messages beyond $2$ has little impact, with
$M_1>3$ even being counterproductive. Intuitively, the congestion caused by many
messages prevents bins from choosing the ``right'' ball to respond to. In the
extreme case of each ball contacting each bin, the situation gets reversed: The
bins ``throw'' $nL_1$ responses ``into $n$ balls'', and the probability for a
ball to not receive a response is $(1-1/n)^{nL_1}\approx e^{-L_1}$.

\textbf{Bin Loads.}
To determine the load distribution after the first round, we compute the
probability $p^{(k)}(M_1,L_1)$ that a given bin gets load
$k\in\{0,\ldots,L_1\}$. To this end, we consider the number of messages $m$ it
received and determine the probability that exactly $k$ out of $\min\{m,L_1\}$
(the number of responses the bin sent) balls will choose this bin. It holds that
\begin{equation}\label{eq:p^k}
p^{(k)}(M_1,L_1)
\approx \sum_{m=k}^{L_1-1}\frac{\E[X_m]}{n}\cdot
\binom{m}{k}p_c^k(1-p_c)^{m-k} 
+\sum_{m=L_1}^{\infty}\frac{\E[X_m]}{n}\cdot \binom{L_1}{k}p_c^k(1-p_c)^{L_1-k},
\end{equation}
where $p_c=p_c(M_1,L_1)$ is the probability that one of the balls contacted by
the bin indeed chooses the bin to commit to. As the ball picks uniformly from
the responding bins, we can instead order the ball's messages' destinations
randomly and pick the first responding bin according to this order. We know that
the considered bin is among them and---up to negligible error---the other
messages will be sent to different bins. Therefore, we can sum over all $M_1$
possible positions of the target bin and multiply $1/M_1$ (the probability that
it is at this position) with the probability that all previous messages do
not receive a response. Writing $p_s=p_s(M_1,L_1)$, we get
\begin{equation}\label{eq:p_c}
p_c\approx\sum_{i=0}^{M_1-1}\frac{(1-p_s)^i}{M_1} =\frac{1-(1-p_s)^{M_1}}{p_s
M_1}.
\end{equation}
\tableref{table:round_1_loads} in the appendix lists, for varying $M_1$, the
derived estimates of $p^{(k)}(M_1,2)$ and $p^{(k)}(M_1,3)$, respectively, and
compares to results from simulations.

\subsection{Ranked Messages}

To avoid the issue that increasing $M_1$ is detrimental, we rank the messages of
each node, and bins give preference to messages of small rank. We can
immediately see that this guarantees that the number of allocated balls must
increase with the number of sent messages, since messages of higher rank do not
affect whether a bin responds to a message of small rank.

We already computed the number of bins receiving a certain number of messages
given $M_1$. We now reuse this information as follows, where $X_m(k)$ denotes
the expected number of bins receiving $m$ messages given that each ball sends
$k$ messages. The probability $p_i(L_1)$ that a message with rank $i\in
\{1,\ldots,M_1\}$ receives a response can be inferred as
\begin{equation*}
p_i(L_1)\approx \sum_{m=0}^{L_1-1}
\frac{\E[X_m(i-1)]}{n}\cdot
\sum_{m'=0}^{\infty}\frac{\E[X_{m'}(1)]}{n}
\cdot\min\left\{\frac{L_1-m}{m'+1},1\right\},
\end{equation*}
where $\E[X_m(i-1)]/n$ is the probability of a bin to receive $m$
messages of rank smaller than $i$, $\E[X_{m'}(1)]/n$ is the probability
to receive $m'$ messages of exactly rank $i$ (different from the considered
message of rank $i$), and $\min\{L_1-m/(m'+1),1\}$ is the probability that a bin
receiving these messages will choose to respond to the ball we consider. Here we
exploit that all respective decisions are made independently and messages of
rank larger than $i$ are of no concern. 

Observe that the inner sum equals $p_s(1,L_1-m)$. Inserting this and the values
for $\E[X_m]$ with $M_1=i-1$ we computed before, we obtain\footnote{We use the
convention that $0^0=1$ here to ensure that the terms are correct for $i=1$ as
well.}
\begin{equation*}
p_i(L_1)\approx \sum_{m=0}^{L_1-1}
\frac{(i-1)^m}{e^{i-1}m!}\,\cdot \, p_s(1,L_1-m).
\end{equation*}
We conclude that the probability $p_{\text{ranked}}(M_1,L_1)$ for a ball to
commit in the first round using ranked messages is
\begin{equation*}
p_{\text{ranked}}(M_1,L_1)=1-\prod_{i=1}^{M_1}(1-p_i(L_1)).
\end{equation*}
Some values together with the results from 100 simulation runs with $10^6$ balls
and bins each are given in \tableref{table:round_1_ranked} in the appendix.

\textbf{Bin Loads.}
Approximating the bin loads algebraically for the algorithm with ranking is
tedious. Since the load is a function of the number of messages of each rank
received, the number of summands increases rapidly with $M_1$. However, the
associated terms decrease exponentially, implying that the number of summands
that need to be considered can be reasonably bounded.

To approximate the probability $p_{\text{ranked}}^{(k)}(M_1,L_1)$ that a bin has
load $k\in \N$ at the end of the first round with ranking, we sum over all vectors
$(k_1,\ldots,k_{M_1}),(m_1,\ldots,m_{M_1})\in \N_0^{M_1}$ that represent
feasible combinations of the number of balls $m_i$ sending a rank $i$ message
to the bin and the number $k_i$ of such balls that commit to the bin due to
a response to such a message, respectively. Hence, the vectors must clearly
satisfy that $m_i\geq k_i$ for all $i$ and that $k=\sum_{i=1}^{M_1}k_i$.
However, it is also necessary that for each $i$ with $k_i\neq 0$, the bin
actually responds to at least $k_i$ messages of rank $i$. This holds true if
(and only if) for each $i$,
$r_i:=\max\{\min\{m_i,L_1-\sum_{j=1}^{i-1}m_j\},0\}\geq k_i$. Out of the $r_i$
balls receiving a response, exactly $k_i$ need to commit to the bin. Overall, we
have that
\begin{equation*}
p_{\text{ranked}}^{(k)}(M_1,L_1)
\approx
\sum_{\substack{(k_1,\ldots,k_{M_1})\in \N_0^{M_1}\\
k=\sum_{i=1}^{M_1}k_i}} \sum_{\substack{
(m_1,\ldots,m_{M_1})\in \N_0^{M_1}\\
\forall i\in\{1,\ldots,M_1\}:\;r_i\geq k_i}}
\prod_{i=1}^{M_1}\frac{\E[X_{m_i}(1)]}{n}\cdot
\binom{r_i}{k_i}\cdot p_c(i)^{k_i}(1-p_c(i))^{r_i-k_i},
\end{equation*}
where
\begin{equation*}
p_c(i)=p_c(i,L_1)=\prod_{j=1}^{i-1}1-p_j(L_1)
\end{equation*}
is the probability that a ball receiving a response to its message of rank $i$
is committing to the respective bin (i.e., it did not receive a response to a
message of smaller rank). Similar to the previous cases, this infinite sum can
be transformed into a finite one, exploiting that we know the limit
$\sum_{m_i=L_i-\sum_{j=1}^{i-1}m_j}^{\infty}\E[X_{m_i}(1)]$. A program can
easily approximately compute $p_{\text{ranked}}^{(k)}(M_1,L_1)$.\footnote{As
$L_i$ and thus $k$ is small, the exponential number of summands in $k$ does not
result in prohibitive complexity.} \tableref{table:round_1_loads_ranked} in the
appendix lists the computed bin loads and compares to the results from simulation.

\section{Later Rounds}\label{sec:later}

Having determined the remaining balls and bin loads at the end of a given round
$r-1$, we can compute these values for round $r$ in a similar fashion as we did
for round $1$; for the considered class of algorithms, no other parameters are
of relevance. We continue to exploit the strong concentration of these random
variables, enabling us to base our computations on expected values without
introducing substantial errors---as long as the number of balls does not become
very small. Once, say $\polylog n$ balls remain, it is likely that all balls
commit in the next round; the computed probability bounds for commitment and
Markov's inequality then yield an estimate of the probability to
terminate.\footnote{Note that our approximation introduces an error of up to
$\polylog n/\sqrt{n}$ in the probability to receive a response to a message,
which is to be taken into account in this bound.}

To simplify the notation, we will use a generic notation with respect to some
variables both for unranked and ranked messages. By $Y_{\ell}$ we denote the
random variable counting the number of bins with load $\ell\in
\{0,\ldots,L_{r-1}\}$ at the end of round $r-1$. Denote by $n_r$ the number
of balls at the beginning of round $r$ and set $\alpha:=M_r\E[n_r]/n$.
The expected number of bins receiving $m$ messages in round $r$ is approximately
\begin{equation*}
\E[X_m]= n\cdot \binom{M_r n_r}{m}
\left(\frac{1}{n}\right)^m\left(1-\frac{1}{n}\right)^{M_r n_r-m}
\approx n\cdot \frac{\alpha^m}{e^{\alpha}m!}.
\end{equation*}
As all random choices are made uniformly and independently, it follows that the
expected number of bins $X_{\ell,m}$ of load $\ell$ that receive $m$ messages is
roughly
$\E[X_{\ell,m}]\approx \alpha^m\E[Y_{\ell}]/(e^{\alpha}m!)$.

\subsection{Unranked Messages}

To estimate the probability $p_s$ that a specific message receives a response in
the unweighted case, we again fix all messages' destinations except for one. Now
we simply sum over all combinations of $\ell$ and $m$, yielding
\begin{equation*}
p_s\approx \sum_{\ell=0}^{L_{r-1}}\sum_{m=0}^{\infty} 
\frac{\E[X_{\ell,m}]}{n}\cdot\min\left\{1,\frac{L_r-\ell}{m+1}\right\}
\approx \sum_{\ell=0}^{L_{r-1}}\frac{\E[Y_{\ell}]}{n}\sum_{m=0}^{\infty} 
\frac{\alpha^m}{e^{\alpha}m!}\cdot\min\left\{1,\frac{L_r-\ell}{m+1}\right\}.
\end{equation*}
Analogously to \eqref{eq:p_round_1}, the probability that a ball
commits is $p\approx 1-(1-p_s)^{M_r}$ and can thus be inferred easily from
$p_s$.

\textbf{Bin Loads.}
To determine the probability $p^{(k)}$ for a bin to have load $k\in
\{0,\ldots,L_r\}$ at the end of round $i+1$, we follow the same approach. We
sum over the loads at the beginning of the round, where each summand is the
probability to have load $\ell\in \{0,\ldots,L-{r-1}\}$ at the beginning of the
round multiplied by the probability to have load $k$ at the end of the round
conditional to this event. Note that having loads $\ell$ and $k$ at the
beginning and end of the round, respectively, is equivalent to being empty,
accepting up $L_r-\ell$ balls, and attaining load $k-\ell$. Using
\eqref{eq:p^k} with these replacements, we obtain
\begin{align*}
p^{(k)}& \approx \sum_{\ell=0}^{L_{r-1}}\frac{\E[Y_{\ell}]}{n}\cdot\left(
\sum_{m=k-\ell}^{L_r-\ell-1}\frac{\E[X_{m}]}{n}\cdot
\binom{m}{k-\ell}p_c^{k-\ell}(1-p_c)^{m-k+\ell}\right.\\ 
&\qquad\qquad\qquad\left.+\sum_{m=L_r-\ell}^{\infty}\frac{\E[X_{m}]}{n}
\cdot\binom{L_r-\ell}{k-\ell}p_c^{k-\ell}(1-p_c)^{L_r-k+\ell}\right),
\end{align*}
where
\begin{equation*}
p_c\approx\sum_{i=0}^{M_r-1}\frac{(1-p_s)^i}{M_r}
=\frac{1-(1-p_s)^{M_r}}{p_s M_r}
\end{equation*}
as in \eqref{eq:p_c}.

\subsection{Ranked Messages}
Applying the same pattern, deriving the expressions for the algorithm with
ranking is now straightforward. We sum over the possible bin loads
$\ell\in \{0,\ldots,L_{r-1}\}$ from the previous round, weigh with the
probability for a bin to have this load, and multiply with the probability for a
bin with (effective) maximal bin load of $L_r-\ell$ in round $r$ to have
$k-\ell$ balls commit to it.
\begin{equation*}
p_i\approx \sum_{\ell=0}^{L_{r-1}}\frac{\E[Y_{\ell}]}{n}\cdot
\sum_{m=0}^{L_r-\ell-1}
\frac{\E[X_{m}(i-1)]}{n}\cdot
\sum_{m'=0}^{\infty}\frac{\E[X_{m'}(1)]}{n}
\cdot\min\left\{\frac{L_r-\ell-m}{m'+1},1\right\},
\end{equation*}
Here $X_m(i-1)$ and $X_{m'}(1)$ denote the variables counting the number of bins
receiving $m$ messages of rank smaller than $i$ and $m'$ messages of rank $i$,
respectively (i.e., $X_m$ when assuming that $i-1$ or $1$ messages are sent
per ball, respectively). As in the first round, the probability for a ball to
commit to some bin then is
\begin{equation*}
p_{\text{ranked}}=1-\prod_{i=1}^{M_r}(1-p_i).
\end{equation*}

\textbf{Bin Loads.}
To determine the bin loads at the end of the round, we need to adjust 
$r_i:=\max\{\min\{m_i,L_r-\ell-\sum_{j=1}^{i-1}m_j\},0\}$, i.e., take into account the bin
load $\ell\in \{0,\ldots,L_{r-1}\}$ carried over from the previous round, and
obtain
\begin{equation}\label{eq:p_k_general}
p_{\text{ranked}}^{(k)}\approx 
\sum_{\ell=0}^{L_{r-1}}\frac{\E[Y_{\ell}]}{n}
\sum_{\substack{(k_1,\ldots,k_{M_r})\in \N_0^{M_r}\\
k-\ell=\sum_{i=1}^{M_r}k_i}} \sum_{\substack{
(m_1,\ldots,m_{M_r})\in \N_0^{M_r}\\
\forall i\in\{1,\ldots,M_r\}:\;r_i\geq k_i}}
\prod_{i=1}^{M_r}\frac{\E[X_{m_i}(1)]}{n}\cdot
\binom{r_i}{k_i}\cdot p_c(i)^{k_i}(1-p_c(i))^{r_i-k_i},
\end{equation}
where
\begin{equation*}
p_c(i)=\prod_{j=1}^{i-1}1-p_j.
\end{equation*}
Apart from $r_i$ and the weighted summation over $\ell\in \{0,\ldots,L_{r-1}\}$,
the only other change here is that $\sum_{i=1}^{M_r}k_i=k-\ell$, since only
$k-\ell$ additional balls need to commit to the bin to reach load $k$.

\subsection{Different Numbers of Balls and Bins}

Note that the expression we gave in this sections can also be applied to the
first round, where we simply have that $Y_0=n$ and $Y_{\ell}=0$ for all $\ell
\neq 0$. Setting $n_1\neq n$ merely changes $\alpha$, without affecting the
expressions in any other way. Thus, our analysis can be applied to the general
case of different numbers of balls and bins.

However, if $n_1\gg n$, $L_i$ will have to be chosen fairly large. This will
render computing the bin loads with ranked messages using \eqref{eq:p_k_general}
problematic, since the number of summands with non-negligible contribution grows
rapidly. This could be tackled by grouping them together into blocks and use
approximate terms (with small error) to simplify the expressions. We note that
$n_1\gg n$ also implies that even the trivial algorithm placing balls uniformly
at random performs well, though, so we refrain from addressing this issue
formally.

\section{Specific Results and Comparison to Other
Algorithms}\label{sec:specific}

Due to the sheer number of possible combinations of the parameters, we believe
that an attempt to discuss the parameter space exhaustively would be fruitless.
Therefore, in this section we discuss several combinations of parameters we
consider of particular interest. To round off the presentation, we make a best
effort at a fair comparison to algorithms from the literature; the relevant
candidates here are variants of the Greedy algorithm~\cite{adler95,even09} and
Stemann's collision algorithm~\cite{stemann96}.

We will focus on choices of parameters that optimize for load, rounds, and the
total number of messages, respectively, while not neglecting the other
optimization criteria. We will constrain the number of bins a ball may contact
in a given round to at most $5$; this more or less arbitrary choice serves to
demonstrate that it is not necessary to enable balls to contact a very large
number of bins concurrently. Given that the performance is strictly better with
ranked messages, we examine only this case. We will keep the bin loads to a
maximum of $2$ or $3$. Under this constraint, \tableref{table:round_1} and
\tableref{table:round_1_ranked} show that there is little to gain in choosing
$M_1>2$. Since a key advantage of adaptiveness is that it permits to keep the
total number of messages small, we will hence keep $M_1\in \{1,2\}$. It turns
out that even with these restrictions, we can do well in $3$ or even $2$ rounds.
Given that $1$ round or maximum load $1$ are clearly insufficient, this leaves a
reasonably small number of options to explore.

The following results have been confirmed by simulations with $10^{6}$ and when
necessary with $10^{7}$ balls to the extent possible; as observed in
\sectionref{sec:preliminaries}, the computed expectations are close to the exact
ones and the respective random variables are strongly concentrated. Spot checks
confirmed that, as expected, standard deviations behave approximately as
$\sqrt{n}$, i.e., for $10^4$ balls the relative deviations from
expectations increase by factor $10$. Given the minuscule variations observed
already for $10^6$ balls, we focus on this number in the following, with the
goal of essentially eliminating one free parameter. We note that in most cases
the expected number of remaining balls at the end of the experiment was far
below $1$, so all balls were placed in the simulations.

\textbf{Minimizing the Maximal Bin Load.}
We use the following set of parameters: $3$ rounds, $L_1=L_2=L_3=2$, $M_1=2$,
$M_2=5$, $M_3=5$. The computed fraction of remaining balls is $5.45\cdot
10^{-7}$.\footnote{Note that if the expected number of remaining balls is
smaller than $1$, Markov's inequality gives a straightforward upper bound on the
probability that not all balls commit. Because we apply Chernoff's bound only in
rounds prior to the last when there are still sufficiently many balls, the
computed expectations are still accurate.} The total number of messages sent is
bounded from above by $n+2R$, where $R$ is the total number of requests sent,
since each request receives a response and each ball sends a final
message to commit. We compute $R\approx 2.23n$, implying that fewer than $5.5n$
messages are sent in total. The fractions of bins with loads $0$, $1$, and $2$
are approximately $31.4\%$, $37.3\%$, and $31.4\%$, respectively.

\textbf{Minimizing the Number of Rounds.}
Here, our goal is to place all balls in two rounds. We choose $M_1=2$ and
$M_2=5$, and pick $L_1=2$ and $L_2=3$. Increasing the permitted load in
the second round has the advantage that all bins still accept at least one ball,
reducing the probability for a ball to have ``collisions'' for all requests. As
a positive side effect, the load distribution improves compared to the case
$L_1=L_2=3$, since fewer bins will have load $3$. An
expected fraction of $5.7\cdot 10^{-10}$ of the balls remain, roughly
$R\approx 2.23n$ messages are sent (i.e., fewer than $5.5n$ total messages),
and the load distribution is about $31.98\%$, $37.37\%$, $29.32\%$, and
$1.33\%$ for loads $0$, $1$, $2$, and $3$, respectively.

\textbf{Minimizing Communication.}
We choose $M_1=1$ and $M_2=M_3=2$. The load sequence is $(2,3,3)$; we note that
compared to the sequence $(3,3,3)$, the expected number of remaining balls drop
by a factor of roughly $250$. The expected fraction of remaining balls is
roughly $4.88\cdot 10^{-8}$, $R\approx 1.21n$ (i.e., fewer than $3.5n$ messages
are sent), and the load distribution is $33.12\%$, $36.60\%$, $27.45\%$, and
$2.83\%$ for loads $0$ to $3$, respectively.

\textbf{Maximizing the Probability to Terminate at Low Communication
Overhead.}
We choose $M_1=1$, $M_2=4$, and $M_3=5$. The load sequence is $(2,2,3)$; we note
that compared to the sequence $(3,3,3)$, the expected number of remaining balls
drops by a factor of roughly $10^7$. The expected fraction of remaining balls is
roughly $5.9\cdot 10^{-19}$, $R\approx 1.41n$ (i.e., fewer than $3.85n$ messages
in total), and the load distribution is $31.759\%$, $36.524\%$, $31.675\%$, and
$0.042\%$ for loads $0$ to $3$, respectively.

\textbf{Comparison to Variants of the Parallel Greedy Algorithm.}
We simulated the simple (``one-shot'') Greedy algorithm~\cite{adler95} for $d=5$
contacted bins and $10^6$ balls and determined the fraction of balls that would
be able to commit if we restricted the bin loads to $2$ and $3$, respectively.
This requires roughly $11n$ messages and a fraction of $1.53\%$ and $2.21\cdot
10^{-4}$ of balls remained, respectively. The message complexity can be reduced
by decreasing the number of bins contacted by each ball, but this would result
in even fewer committing balls.

For the multi-round version of Greedy~\cite{adler95} with $d=5$ and $n=10^7$, we
determined the fraction of balls that could be placed in $3$ rounds (resulting
in maximal load $3$). This also resulted in roughly $11n$ messages; after 2
rounds, a fraction $1.14\%$ of the balls remained, while all balls were
placed in 3 rounds. In comparison, our algorithm with $L_1=2$, $L_2=3$, $M_1=1$,
and $M_2=2$ retains a fraction of $6.1\cdot 10^{-5}$ of the balls after 2
rounds, i.e., performs notably better at lower communication complexity.

In~\cite{even09}, the authors propose an adaptive variant of the multi-round
Greedy algorithm called H-{\scshape retry} that runs for 3 rounds. After running
an initial round of the multi-round Greedy algorithm with $d=2$ in the first
round and trying to resolve conflicts in the second round, balls that are still
unsuccessful contact 2 additional bins in the third round. The authors report
simulation results. These indicate that the fraction of remaining balls after 3
rounds is slightly above $10^{-7}$ for bin loads of $3$; the number of messages
is larger than $5n$. This is outperformed in all considered criteria by our
algorithm for message and load sequences $(1,2,2)$ and $(2,3,3)$, respectively.

In summary, we see that the Greedy algorithm compares unfavorably to our
approach, even if we permit each ball to contact $4$ or $5$ different bins and
send a substantially larger number of messages.

\textbf{Comparison to Stemann's Collision Algorithm.}
We ran Stemann's algorithm with accepted loads of $2$ and $3$, respectively. In
Stemann's algorithm, each ball contacts 2 bins. In each round, the bins for
which the accepted load threshold $L$ is large enough to accommodate all
uncommitted balls that contacted them initially inform the respective balls,
which then commit to (one of) the accepting bin(s). This process can be
implemented by each ball (i) sending the initial requests, (ii) sending a commit
message to the respective bin, and (iii) sending a ``will not commit'' message
to the bin it initially contacted but does not commit to. Since there are $2n$
initial requests only, the total number of messages sent by bins will be at
most $2n$. This results in a total of at most $6n$ messages. Roughly $n$ of
these messages can be saved because balls do not need to send a ``will not
commit'' message in case both of the bins they contacted accept them in the same
round, and bins do not need to inform a ball that committed in an earlier round
that it could be accepted.

For load $2$ and $n=10^7$, after $3$ rounds the fraction of remaining balls is
$2.09\%$, and $4.94n$ messages have been sent for the implementation described
above. For load $3$, after $2$ rounds a fraction of $7.8\cdot 10^{-4}$ of the
balls remained and $4.998n$ messages had been sent. In round $3$ the remaining
balls all committed and the message total increased to $5n$.

In comparison, for the parameters $M_1=1$, $M_2=M_3=2$, $L_1=2$, and $L_2=L_3=3$
we picked to minimize communication, after two rounds the fraction of remaining
balls was $6.1\cdot 10^{-5}$; recall that the total number of messages sent was
smaller than $3.5n$. We conclude that even under the constraint that balls
send no more than $2$ messages in each round, our approach outperforms Stemann's
algorithm in terms of the achievable trade-off between maximal load and
communication. Moreover, since in Stemann's algorithm loads of $L$ are accepted
right from the start, for $L=3$ a fraction of $5.51\%$ of the bins ended up with
load $3$, whereas in our case only $2.83\%$ of the bins had this load.

\section{Conclusion}\label{sec:conclusion}

We presented a novel class of simple adaptive algorithms and an accompanying
analysis technique for the parallel balls-into-bins problem. Analytical and
experimental results show substantial improvements over previous algorithms. We
hope that this work and the accompanying simulation code~\cite{scripts} provide
tools for practitioners looking for a distributed balls-into-bins routine
tailored to their needs.

In this paper, we restricted our attention to the synchronous setting. However,
we believe that the presented approach bears promise also for asynchronous
systems. If bins process messages in the order of their arrival and message
delays are independently and uniformly distributed, the resulting behavior of
the algorithm would be identical if no messages from round $i+1$ arrive at a bin
before all messages from round $i$ are processed. To handle this case, a bin can
delay processing messages from rounds larger than $i$ until it is not expecting
a response from a ball which it permitted to commit to it anymore. If a message
from a later round is processed by a bin not awaiting any further responses, we
argue that it is actually beneficial to favor the request over those of earlier
rounds, since the respective ball is in greater need to commit to a bin. This
reasoning suggests that the respective asynchronous variants of our algorithms
provide promising heuristics for asynchronous systems; it also seems plausible
that our analysis technique can be extended to establish worst-case bounds under
asynchrony. Such hope does not exist for algorithms from the literature with low
communication overhead, like H-{\scshape retry} or Stemann's collision
algorithm, whose strategies cannot work without synchronization points.

\bibliographystyle{abbrv}
\bibliography{../balls}

\appendix

\section*{Appendix}
\vspace*{-.05cm}

\begin{table}[h!]
\begin{minipage}{\textwidth}
\begin{center}
\begin{tabular}{c|c|c|c}
$M_1$ & estimated \% & avg.\ \%  & max.\ \%\\\hline
1 & 10.364 & 10.364 & 10.550\\
2 & 7.333 & 7.346 & 7.380\\
3 & 7.222 & 7.218 & 7.412\\
4 & 7.774 & 7.740 & 7.870\\
5 &  8.407 & 8.413 & 8.466\\
10 & 10.745 & 10.732 & 10.860\\
20 & 12.158 & 12.177 & 12.252\\
$\infty$ & 13.536 & &
\end{tabular}
\qquad
\begin{tabular}{c|c|c|c}
$M_1$ & estimated \% & avg.\ \%  & max.\ \%\\\hline
1 & 2.334 & 2.333 & 2.465\\
2 & 1.188 & 1.182 & 1.249\\
3 & 1.125 & 1.131 &  1.228\\
4 & 1.290 & 1.288 & 1.400\\
5 &  1.546 & 1.549 & 1.678\\
10 & 2.838 & 2.848 & 2.981\\
20 & 3.876 & 3.878 & 4.007\\
$\infty$ & 4.978 & &
\end{tabular}
\end{center}
\end{minipage}
\caption{Remaining balls after one round, for $L_1=2$ (left) and
$L_1=3$ (right), without ranks. 100 simulation runs were performed with $10^6$
balls each. The entry ``$\infty$'' gives the limit for $M_1\to
\infty$.}\label{table:round_1}
\end{table}
\vspace*{-.05cm}

\begin{table}[h!]
\small
\begin{minipage}{\textwidth}
\begin{center}
\begin{tabular}{c|c|c|c|c}
$M_1$ & load & estimated \% & avg.\ \%  & max.\ \%\\\hline
\multirow{3}{*}{1} & 0 & 36.788 & 36.785 & 37.180\\
& 1 & 36.788 & 36.792 & 37.143\\
& 2 & 26.424 & 26.423 & 26.766\\
& & & & \\\hline
\multirow{3}{*}{2} & 0 & 31.303 & 31.310 & 31.512\\
& 1 & 44.720 & 44.710 & 45.098\\
& 2 & 23.977 & 23.981 & 24.167\\
& & & & \\\hline
\multirow{3}{*}{3} & 0 & 29.701 & 29.715 & 29.989\\
& 1 & 47.814 & 47.791 & 48.221\\
& 2 & 22.485 & 22.494 & 22.671\\
& & & & \\\hline
\multirow{3}{*}{4} & 0 & 29.384 & 29.385 & 29.566\\
& 1 & 48.971 & 48.976 & 49.269\\
& 2 & 21.644 & 21.639 & 21.845\\
& & & & \\\hline
\multirow{3}{*}{5} & 0 & 29.516 & 29.521 & 29.712\\
& 1 & 49.375 & 49.366 & 49.736\\
& 2 & 21.109 & 21.112 & 21.314\\
& & & & \\\hline
\multirow{3}{*}{10} & 0 & 30.662 & 30.662 & 30.992\\
& 1 & 49.421 & 49.414 & 49.767\\
& 2 & 19.917 & 19.923 & 20.094\\
& & & & \\\hline
\multirow{3}{*}{20} & 0 & 31.448 & 31.454 & 31.647\\
& 1 & 49.261 & 49.251 & 49.692\\
& 2 & 19.291 & 19.295 & 19.469\\
& & & & \\\hline
\multirow{3}{*}{$\infty$} & 0 & 32.203 & & \\
& 1 & 49.089 & & \\
& 2 & 18.708 & & \\
& & & &
\end{tabular}
\qquad
\begin{tabular}{c|c|c|c|c}
$M_1$ & load & estimated \% & avg.\ \%  & max.\ \%\\\hline
\multirow{4}{*}{1} & 0 & 36.788 & 36.779 & 36.990\\
& 1 & 36.788 & 36.807 & 37.155\\
& 2 & 18.394 & 18.386 & 18.707\\
& 3 & 8.030 & 8.029 & 8.214\\\hline
\multirow{4}{*}{2} & 0 & 33.822 & 33.814 & 34.151\\
& 1 & 39.056 & 39.067 & 39.473\\
& 2 & 21.609 & 21.606 & 21.877\\
& 3 & 5.513 & 5.513 & 5.635\\\hline
\multirow{4}{*}{3} & 0 & 32.439 & 31.968 & 32.208\\
& 1 & 42.664 & 41.615 & 42.013\\
& 2 & 22.776 & 21.998 & 22.282\\
& 3 & 4.611 & 4.419 & 4.548\\\hline
\multirow{4}{*}{4} & 0 & 31.057 & 31.055 & 31.250\\
& 1 & 43.105 & 43.122 & 43.584\\
& 2 & 21.910 & 21.881 & 22.174\\
& 3 & 3.993 & 3.942 & 4.037\\\hline
\multirow{4}{*}{5} & 0 & 30.791 & 30.696 & 30.913\\
& 1 & 43.993 & 43.848 & 44.330\\
& 2 & 21.846 & 21.761 & 21.995\\
& 3 & 3.707 & 3.695 & 3.810\\\hline
\multirow{4}{*}{10} & 0 & 30.913 & 30.919 & 31142\\
& 1 & 44.411 & 44.395 & 44.762\\
& 2 & 21.277 & 21.289 & 21.560\\
& 3 & 3.399 & 3.396 & 3.483\\\hline
\multirow{4}{*}{20} & 0 & 31.386 & 31.383 & 31.600\\
& 1 & 44.394 & 44.391 & 44.932\\
& 2 & 20.931 & 20.937 & 21.280\\
& 3 & 3.290 & 3.289 & 3.443\\\hline
\multirow{4}{*}{$\infty$} & 0 & 31.883 & & \\
& 1 & 44.362 & & \\
& 2 & 20.575 & & \\
& 3 & 3.181 & & \\
\end{tabular}
\end{center}
\end{minipage}
\caption{Fractions of bins with a given load after round one, for $L_1=2$ (left)
and $L_1=3$ (right), without ranks and 100 simulation runs with $n=10^6$. Entry
``$\infty$'' gives the limit for $M_1\to \infty$.}\label{table:round_1_loads}
\end{table}

\begin{table}[pb]
\begin{minipage}{\textwidth}
\begin{center}
\begin{tabular}{c|c|c|c}
$M_1$ & estimated \% & avg.\ \%  & max.\ \%\\\hline
1 & 10.364 & 10.372 & 10.622\\
2 & 4.536 & 4.542 & 4.703\\
3 & 3.210 & 3.212 & 3.323\\
4 & 2.764 & 2.760 & 2.858\\
5 &  2.590 & 2.593 & 2.734\\
10 & 2.471 & 2.471 & 2.608\\
20 & 2.470 & 2.474 & 2.589\\
$\infty$ & 2.470 & &
\end{tabular}
\qquad
\begin{tabular}{c|c|c|c}
$M_1$ & estimated \% & avg.\ \%  & max.\ \%\\\hline
1 & 2.334 & 2.340 & 2.434\\
2 & 0.454 & 0.455 & 0.510\\
3 & 0.206 & 0.205 & 0.234\\
4 & 0.139 & 0.139 & 0.170\\
5 &  0.115 & 0.115 & 0.138\\
10 & 0.097 & 0.984 & 0.123\\
20 & 0.096 & 0.974 & 0.123 \\
$\infty$ & 0.096 & &
\end{tabular}
\end{center}
\end{minipage}
\caption{Remaining balls at the end of the first round with ranking, for $L_1=2$
(left) and $L_1=3$ (right). 100 simulation runs were performed with $10^6$ balls
each. The entry ``$\infty$'' gives the limit for $M_1\to
\infty$.}\label{table:round_1_ranked}
\end{table}

\begin{table}[p]
\small
\begin{minipage}{\textwidth}
\begin{center}
\begin{tabular}{c|c|c|c|c}
$M_1$ & load & estimated \% & avg.\ \%  & max.\ \%\\\hline
\multirow{3}{*}{1} & 0 & 36.788 & 36.794 & 37.014\\
& 1 & 36.788 & 36.771 & 37.203\\
& 2 & 26.424 & 26.435 & 26.591\\
& & & & \\\hline
\multirow{3}{*}{2} & 0 & 33.475 & 33.484 & 33.739\\
& 1 & 37.585 & 35.576 & 37.881\\
& 2 & 28.939 & 28.940 & 29.128\\
& & & & \\\hline
\multirow{3}{*}{3} & 0 & 32.584 & 32.578 & 32.816\\
& 1 & 38.042 & 38.045 & 38.415\\
& 2 & 29.374 & 29.377 & 29.572\\
& & & & \\\hline
\multirow{3}{*}{4} & 0 & 32.255 & 32.239 & 32.408\\
& 1 & 38.253 & 38.279 & 38.603\\
& 2 & 29.492 & 29.482 & 29.660\\
& & & & \\\hline
\multirow{3}{*}{5} & 0 & 32.112 & 32.112 & 32.300\\
& 1 & 38.350 & 38.357 & 38.726\\
& 2 & 29.530 & 29.531 & 29.700\\
& & & & \\\hline
\multirow{3}{*}{10} & 0 & 32.022 & 32.026 & 32.265\\
& 1 & 38.427 & 38.418 & 38.770\\
& 2 & 29.551 & 29.556 & 29.696\\
& & & & \\\hline
\multirow{3}{*}{20} & 0 & 32.021 & 32.037 & 32.245\\
& 1 & 38.428 & 38.394 & 38.822\\
& 2 & 29.551 & 29.569 & 29.732\\
& & & &
\end{tabular}
\qquad
\begin{tabular}{c|c|c|c|c}
$M_1$ & load & estimated \% & avg.\ \%  & max.\ \%\\\hline
\multirow{4}{*}{1} & 0 & 36.788 & 36.783 & 37.000\\
& 1 & 36.788 & 36.792 & 37.238\\
& 2 & 18.394 & 18.392 & 18.639\\
& 3 & 8.030 & 8.032 & 8.195\\\hline
\multirow{4}{*}{2} & 0 & 35.958 & 35.957 & 36.256\\
& 1 & 36.845 & 36.843 & 37.179\\
& 2 & 18.890 & 18.898 & 19.200\\
& 3 & 8.307 & 8.301 & 8.501\\\hline
\multirow{4}{*}{3} & 0 & 35.826 & 35.820 & 36.041\\
& 1 & 36.882 & 36.891 & 37.299\\
& 2 & 18.965 & 18.965 & 19.229\\
& 3 & 8.327 & 8.324 & 8.499\\\hline
\multirow{4}{*}{4} & 0 & 35.785 & 35.790 & 36.064\\
& 1 & 36.900 & 36.894 & 37.226\\
& 2 & 18.984 & 18.982 & 19.292\\
& 3 & 8.330 & 8.335 & 8.484\\\hline
\multirow{4}{*}{5} & 0 & 36.769 & 35.778 & 36.029\\
& 1 & 36.909 & 36.903 & 37.306\\
& 2 & 18.991 & 18.973 & 19.176\\
& 3 & 8.332 & 8.346 & 8.487\\\hline
\multirow{4}{*}{10} & 0 & 35.755 & 35.747 & 35.949\\
& 1 & 36.918 & 36.934 & 37.295\\
& 2 & 18.995 & 18.986 & 19.245\\
& 3 & 8.332 & 8.333 & 8.494\\\hline
\multirow{4}{*}{20} & 0 & 37.755 & 35.743 & 35.939\\
& 1 & 36.919 & 36.935 & 37.299\\
& 2 & 18.995 & 18.996 & 19.218\\
& 3 & 8.332 & 8.326 & 8.541
\end{tabular}
\end{center}
\end{minipage}
\caption{Percentage of bins with a given load, for $L_1=2$ (left) and
$L_1=3$ (right), with ranks. 100 simulation runs were performed with $10^6$
balls each.}\label{table:round_1_loads_ranked}
\end{table}

\end{document}